\RequirePackage[final]{graphicx}
\documentclass{new_tlp}

\usepackage{amsmath}
\usepackage{amssymb}
\usepackage{graphicx}
\usepackage{enumitem}

\usepackage[multiple]{footmisc}

\newtheorem{thm}{Theorem}[section]

\newcommand{\etal}{~et~al.}

\newcommand{\tuple}[1]{\langle{#1}\rangle}
\newcommand{\fix}{\lfp}

\newcommand{\PH}{{\powerset {H_P}}}
\newcommand{\interp}[1]{[{#1}]}
\newcommand{\T}{\widehat T}

\newcommand{\Lcurly}{{L(\preccurlyeq)}}

\newcommand{\BProlog}[0]{\mbox{B-Prolog}}

\newcommand{\tabledomain}{I_P \times U_P^n \to U_P^\bot}

\title{Tabling with Sound Answer Subsumption}

\author[Vandenbroucke\etal]{
  Alexander Vandenbroucke\\
  KU Leuven, Belgium\\
  \email{alexander.vandenbroucke@kuleuven.be}
  \and
  Maciej Pir{\'o}g\\
  KU Leuven, Belgium\\
  \email{maciej.pirog@kuleuven.be}
  \and
  Benoit Desouter\\
  Ghent University, Belgium\\
  \email{benoit.desouter@ugent.be}
  \and
  Tom Schrijvers\\
  KU Leuven, Belgium\\
  \email{tom.schrijvers@kuleuven.be}
}

\submitted{29 April 2016}
\revised{8 July 2016}
\accepted{22 July 2016}
\pagerange{\pageref{firstpage}--\pageref{lastpage}}

\usepackage{ifdraft}
\usepackage{todonotes}

\ifdraft{
\newcommand{\benoit}[1]{\todo{B: #1}}
}{
\newcommand{\benoit}[1]{}
}

\ifdraft{
\newcommand{\alexander}[1]{\todo{A: #1}}
}{
\newcommand{\alexander}[1]{}
}

\ifdraft{
\newcommand{\tom}[1]{\todo{T: #1}}
}{
\newcommand{\tom}[1]{}
}

\ifdraft{
\newcommand{\maciej}[1]{\todo{M: #1}}
}{
\newcommand{\maciej}[1]{}
}

\usepackage{amsmath}
\usepackage{amssymb}
\usepackage{fancyvrb}
\usepackage{float}
\usepackage{relsize}
\usepackage{booktabs}
\usepackage{bm}
\usepackage{url}

\usepackage{tikz}
\usetikzlibrary{positioning}

\newcounter{ExampleCounter}
\newenvironment{example}[0]{\refstepcounter{ExampleCounter}\noindent\paragraph{Example~\theExampleCounter}}{}

\DefineVerbatimEnvironment{CodeVerbatim}{Verbatim}{fontsize=\footnotesize}

 \newcommand{\lfp}{\mathsf{lfp}}

\newcommand{\vect}[1]{\ensuremath{\boldsymbol{#1}}} 

\newcommand{\powerset}[1]{\ensuremath{\mathcal{P}\left(#1\right)}}
\newcommand{\powersetfin}[1]{\ensuremath{\mathcal{P}^{\mathsf{fin}}\left(#1\right)}}

\newcommand{\emptySetOfBaseType}[1]{\ensuremath{\varnothing}}

\begin{document}

\label{firstpage}
\maketitle


\begin{abstract}
Tabling is a powerful resolution mechanism for logic programs that 
captures their least fixed point semantics more faithfully than plain Prolog. In many
tabling applications, we are not interested in the set of all answers to a
goal, but only require an aggregation of those answers. Several works have
studied efficient techniques, such as lattice-based answer subsumption and
mode-directed tabling, to do so for various forms of aggregation.

While much attention has been paid to expressivity and efficient implementation
of the different approaches, soundness has not been considered. This paper shows
that the different implementations indeed fail to produce least fixed points for
some programs. As a remedy, we provide a formal framework that generalises the
existing approaches and we establish a soundness criterion that explains for
which programs the approach is sound.

\noindent
\emph{This article is under consideration for acceptance in TPLP.}
\end{abstract}



\section{Introduction}
Tabling considerably improves the declarativity and expressiveness of the
Prolog language. It removes the sensitivity of SLD resolution to rule and goal
ordering, allowing a larger class of programs to terminate.  As an added bonus,
the memoisation of the tabling mechanism may significantly improve run time
performance in exchange for increased memory usage.  Tabling has been
implemented in a few well-known Prolog systems, such as
XSB~\cite{SwiW10a,swift2012},Yap~\cite{yap}, Ciao~\cite{guzman2008} and
\BProlog{}~\cite{bprolog}, and has been successfully applied in various domains.

Much research effort has been devoted to improving the performance of
tabling for various specialised
use-cases~\cite{swift1999,Ramakrishna1997,zhouSokoban2011}. This paper
is concerned with one such fairly broad class of use-cases: we are not
directly interested in all the answers to a tabled-predicate query,
but instead wish to aggregate these answers somehow.  The following
shortest-path example illustrates this use-case.
\begin{CodeVerbatim}
   query(X,Y,MinDist) :- findall(Dist,p(X,Y,Dist),List), min_list(List,MinDist).
   :- table p/3.
   p(X,Y,1) :- e(X,Y).
   p(X,Y,D) :- p(X,Z,D1), p(Z,Y,D2), D is D1 + D2.
   e(a,b). e(b,c). e(a,c).
\end{CodeVerbatim}
The query \texttt{?- query(a,c,D).} computes the distance \texttt{D} of
the shortest path from \texttt{a} to \texttt{c} by first computing
the set of distances $\{1, 2\}$ of all paths and then selecting the smallest
value from this set. Unfortunately, when the graph is cyclic, the
set of distances is infinite and the query never returns, even though 
the infinite set has a well-defined minimal value.

Various tabling extensions (know collectively as answer subsumption:
mode-directed
tabling~\cite{guo2004mode,guo2008modeExpanded,zhou2010mode,santos2013mode},
partial order answer subsumption and lattice answer
subsumption~\cite{swift2012}), have come up with ways to integrate the
aggregation into the tabled resolution.  This way answers are
incrementally aggregated and the tabling may converge more quickly to
the desired results.  For instance, the shortest-path program can be
written with mode-directed tabling as:
\begin{CodeVerbatim}
  :- table p(+,+,min).
  p(X,Y,1) :- e(X,Y).
  p(X,Y,D) :- p(X,Z,D1), p(Z,Y,D2), D is D1 + D2.
  e(a,b). e(b,c). e(a,c).
\end{CodeVerbatim}
Here the query \texttt{?- p(a,c,D).} yields only the shortest distance.  It
does so by greedily throwing away non-optimal intermediate results and in this
way only considers finitely many paths, even if the graph is cyclic. In
summary, this approach makes tabling (sometimes infinitely) more efficient for
our aggregation use-case.

Unfortunately, none of the existing implementations that we are aware of 
is generally sound. Consider the following pure logic program.
\begin{CodeVerbatim}
   p(0). p(1).
   p(2) :- p(X), X = 1.
   p(3) :- p(X), X = 0.
\end{CodeVerbatim}
The query \texttt{?- p(X).} has a finite set of solutions,
$\{\texttt{p(0),p(1),p(2),p(3)}\}$, the largest of which is
\texttt{p(3)}.  However XSB, Yap and \BProlog{} all yield different
(invalid) solutions when answer subsumption is used to obtain the
maximal value.  Both XSB and \BProlog{} yield \texttt{X = 2}, with a
maximum lattice and \texttt{max} table mode respectively.  Yap (also
with \texttt{max} table mode) yields \texttt{X = 0; X = 1; X = 2},
every solution except the right one.\footnote{The batch scheduling
  used by Yap returns any answer as soon as it is found.}
  
Clearly, these results are unsound. This example is not the only
erroneous one; we can easily construct more erroneous scenarios with
other supported forms of aggregation.  Hence, we must conclude that
answer subsumption is in general not a semantics-preserving
optimisation. Yet, as far as we know, the existing literature does not
offer any guidance on when the feature can be relied upon.  In fact,
to our knowledge, its semantics have not been formally discussed
before.

%
%
%
%

This paper fills the semantic gap of answer subsumption with the help
of lattice theory. We show how the existing implementations fit into
this semantic framework and derive a sufficient condition for
semantics preservation that allows answer subsumption to be safely
used.

\section{Background: Tabling Semantics}
\label{sec:denotationalSemanticsAnswerTabling}

\newcommand{\ground}{\mathsf{ground}}

Because the operational semantics of tabling is rather complex and
different systems vary in subtle ways, we make a simplifying
assumption and assume that tabling systems implement Lloyd's
least fixed-point semantics~\cite{Lloyd:1984} for definite
logic programs, that is, that tabling is a sound program optimisation
with no impact on the denotation of a program. This semantics
conveniently abstracts from low-level aspects such as clause and goal
ordering and the specific clause scheduling algorithm used by the
Prolog engine.

\subsection{Least Fixed-Point Semantics}

First, we need the notion of a \emph{Herbrand base}: the set of all possible
(ground) atoms that occur in a logic program. More formally, let $\Sigma$ be an
alphabet, and $P$ be a logic program over $\Sigma$, then the Herbrand base $H_P$
is the set of all ground atoms over $\Sigma$. For example, the Herbrand
base of the shortest path program (without \texttt{query/3},\texttt{+/2}) is:
\[
H_P = \{\texttt{e(}X,Y\texttt{)}~|~X,Y \in \{\texttt{a},\texttt{b},\texttt{c}\}\}
\cup \{\texttt{p(}X,Y,D\texttt{)}~|~X,Y \in \{\texttt{a},\texttt{b},\texttt{c}\},
D \in \mathbb{N}\}
\]
A \emph{(Herbrand) interpretation} is a set
$I \subseteq H_P$. Intuitively, it contains atoms in the Herbrand base that are true:
$\forall a \in H_P: I \models a \iff a \in I$.

Finally, define the operator $T_P:\PH \to \PH$ such that, given an interpretation~$I$, the value $T_P(I)$ is the interpretation that immediately follows from $I$ by
any of the program rules:
\begin{equation}
\label{eq:Tp}
  T_P(I) = \{
  B_0 \in H_P \mid B_0 \leftarrow B_1, \ldots, B_n \in \ground(P)
  \wedge \{ B_1, \ldots, B_n \} \subseteq I \}
\end{equation}
This operator is called the \emph{immediate consequence operator}.
Its least fixed-point with respect to subset-inclusion ($\subseteq$), denoted
$\lfp(T_P)$, defines the semantics of the program~$P$, and
is also known as the \emph{least Herbrand model}.
It is the interpretation that contains those and only those atoms that
follow from the program and that are not self-supported.

\begin{example}
Consider the following program $P$:
\begin{Verbatim}[fontsize=\footnotesize]
p(a). p(b). q(c).
q(X) :- p(X).
\end{Verbatim}
Its Herbrand base is $\{\texttt{p(a)},\texttt{p(b)},\texttt{p(c)},
\texttt{q(a)},\texttt{q(b)},\texttt{q(c)}\}$. Its fixed-point semantics is:
$$\fix (T_P) = \{ \texttt{p(a)}, \texttt{p(b)}, \texttt{q(a)}, \texttt{q(b)},
\texttt{q(c)} \}$$
Observe that this is exactly the set of atoms that follow from the program.
\end{example}

\subsection{Existence and Computability of the Least Herbrand Model}

The least fixed-point semantics is not necessarily well-defined: it is
not immediate that the least fixed-point actually exists. Moreover, if
it exists, it may not actually be constructively computable.

Fortunately, there is no reason for concern: by appeal to a well-known theorem
from lattice theory, we can easily establish the well-definedness.
A \emph{complete lattice} is a partially ordered set (poset)
$\tuple{L,\leq_L}$ such that every $X \subseteq L$ has a least
upper bound~$\bigvee X$, i.e.:
\[\forall z \in L: \bigvee X \leq_L z \iff \forall x \in X: x \leq_L z\]
We do indeed have a lattice structure at hand: the power set of the Herbrand
base $\tuple{\PH,\subseteq}$ is a complete lattice. In fact, any power set
is a complete lattice.
Moreover, it is quite easy to see that if $P$ is a definite logic program
(i.e., contains no negations), then $T_P$ is monotone with respect to this
lattice.
It follows that $\fix (T_P)$ exists, ensuring that the semantics is
well-defined for every definite program $P$, by the following theorem:
\begin{thm}[Knaster--Tarski]
Let $\tuple{L,\leq_L}$ be a complete lattice, and let $f : L \to L$ be a
monotone function. Then, $f$ has a least fixed point, denoted $\fix(f)$.
\end{thm}

Moreover, the $T_P$ operator is $\omega$-continuous, which means that for
all ascending chains $l_1 \subseteq l_2 \subseteq \ldots$ with
$l_1,l_2, \ldots \subseteq H_P$, it is the case that
$\bigcup_{i=1}^\infty T_P(l_i) = T_P(\bigcup_{i=0}^\infty l_i)$.
Then, Kleene's fixed-point theorem gives us a constructive way of
obtaining $\fix (T_P)$:
 \[ \fix (T_P) = \bigcup \{T_P(\emptyset),T_P^2(\emptyset),\ldots\} \]

The least fixed-point can therefore be obtained in a bottom-up
fashion by iterating $T_P$ from the empty set onward. Operationally,
tabling usually interleaves a top-down goal-directed strategy with
bottom-up iteration. The bottom-up strategy always computes the
entire least Herbrand model, even when only a small portion of it may
be required to prove a particular query.  The top-down part of
tabling avoids computing irrelevant atoms as much as possible, making
inference feasible.
 
\subsection{Stratification}
\label{sec:stratification}

Unfortunately,  the $T_P$-operator is not monotone for programs containing
more advanced constructs, such as negation.
Therefore Lloyd's semantics as described above is not suitable for capturing
the semantics of such programs.
In the case of negation, this problem is solved by partitioning the clauses
of a program into an ordered set of \emph{strata} based on their
interdependence.
This procedure is called \emph{stratification}~\cite{apt1988towards}.
Then, the semantics for each stratum is computed based on the semantics of
the lower strata, with no relation to the higher strata.
To make this more concrete, suppose a ground program $P$ admits a
stratification $P_1,\ldots,P_n$, with the $P_i$ non-empty and pairwise disjoint,
then:
\begin{align*}
  &P = P_1 \cup \cdots \cup P_n,\\
  &Q_1 = P_1,~Q_{i+1} = P_{i+1} \cup M_i,  &&\text{ for all } i=0,\ldots,n-1\\
  &M_i = \fix(T_{Q_i})                   &&\text{ for all } i = 1,\ldots n
\end{align*}
where $M_i$ should be understood as a set of facts.
If a program admits a stratification where all negated calls are to
predicates defined in lower strata, the obvious extension $T_P^{neg}$ of the
$T_P$ operator to include negation is guaranteed to be monotone.
The semantics of $P$ is then given by $\bigcup_{i = 1}^n M_i$.

\section{Answer Subsumption Approaches}
\label{sec:denotationalSemanticsAnswerSubsumption}

In this section, we propose a denotational semantics for tabling with
answer subsumption. Broadly speaking, we modify the semantics for
stratified programs as described in the previous section in two
respects. First, our semantics includes new answers that may emerge
from the program-defined rules of subsumption, which are not
necessarily logical consequences of the same program without answer
subsumption. We obtain this by extending the $T_P$
operator. Secondly, we perform the actual subsumption, that is, we
remove the subsumed answers. Stratification, as discussed in
Section~\ref{sec:latticeStratified}, is used to control the order in
which these two steps are invoked.

In a bit more detail, the semantics of a stratum is given by the
extended immediate consequence operator, which we call $\T_P$, and a
function $\eta : H_P \rightarrow L$ that aggregates the answers using
a lattice $L$.  A consequence of this specification is that an
aggregation naturally ignores all operational aspects of the program
P.  That is to say, given two structurally distinct programs $P_1$ and
$P_2$ whose least fixed-point semantics coincide, that is, $\fix(\T_{P_1})
= \fix(\T_{P_2})$, it follows that $\bigvee\eta(\fix(\T_{P_1})) =
\bigvee\eta(\fix(\T_{P_2}))$, i.e. their aggregates coincide as well.

Obviously, the existing systems do not implement answer subsumption as
a single post-processing function. Instead, they execute it repeatedly
during the bottom-up phase of the computation, which sometimes makes
them deviate from the intended semantics, as exemplified in the
introduction. We formalise and deal with this in Section~\ref{sec:fusion}.

For now, we assume that the program~$P$ has only one stratum.
Towards the end of the section, we show how to assemble the semantics of
programs with any number of strata.

\subsection{Mode-Directed Tabling}
\label{sec:mode-directed-tabling}
Mode-directed tabling is a convenient aggregation approach supported by
ALS-Prolog, \BProlog{} and Yap where the arguments of a tabled
predicate are annotated with one of a range of aggregation \emph{modes}. Yap
provides the largest range of possible modes: \texttt{index}, \texttt{first},
\texttt{last}, \texttt{min}, \texttt{max} and \texttt{all}.
The answers are grouped by distinct values for the \texttt{index} arguments;
the remaining arguments are aggregated according to their mode -- the mode
names should be self-explanatory.

Based on our assumptions so far, we can immediately disqualify the
existing implementations of the three modes \texttt{first}, \texttt{last} and
\texttt{sum}. The reason is that the semantics of the existing implementations
is inherently sensitive to the program structure.
Consider the two programs below:
\vspace{-4mm}
\begin{figure}[H]
\begin{minipage}[t]{0.49\textwidth}
\begin{CodeVerbatim}
  % P1
  :- table p(first).
  p(1). p(2).
\end{CodeVerbatim}
\end{minipage}
\vspace{-4mm}
\begin{minipage}[t]{0.49\textwidth}
\begin{CodeVerbatim}
  % P2
  :- table p(first).
  p(2). p(1).
\end{CodeVerbatim}
\end{minipage}
\end{figure}
Clearly $\fix (T_{P1}) = \{\texttt{p(1)},\texttt{p(2)}\} = \fix (T_{P2})$,
however $P1$ yields
\texttt{p(1)} as an answer for \texttt{?- p(X).} while $P2$ yields
\texttt{p(2)}.
The opposite happens with the \texttt{last} mode.
The next programs illustrate the problem of the \texttt{sum} mode:
\vspace{-4mm}
\begin{figure}[H]
\begin{minipage}[t]{0.49\textwidth}
\begin{CodeVerbatim}
  % P3
  :- table p(sum).
  p(1).
\end{CodeVerbatim}
\end{minipage}
\begin{minipage}[t]{0.49\textwidth}
\begin{CodeVerbatim}
  % P4
  :- table p(sum).
  p(1). p(1).
\end{CodeVerbatim}
\end{minipage}
\end{figure}
Again the least fixed-point semantics of both programs coincides: $\fix
(T_{P3}) = \{\texttt{p(1)}\} = \fix (T_{P4})$.
However they produce the following results in Yap:
\vspace{-4mm}
\begin{figure}[H]
\begin{minipage}[t]{0.49\textwidth}
\begin{CodeVerbatim}
  % P3
  ?- p(X).
  X = 1.
\end{CodeVerbatim}
\end{minipage}
\begin{minipage}[t]{0.49\textwidth}
\begin{CodeVerbatim}
  % P4
  ?- p(X).
  X = 1 ; X = 1.
  ?- p(X).
  X = 2.
\end{CodeVerbatim}
\end{minipage}
\end{figure}
\vspace{-4mm}
Yap produces the result \texttt{p(1)} twice the first time the query is called.
Any subsequent query is answered with \texttt{p(2)}.
In other words, not only are the results of $P3$ and $P4$ not consistent, the
results for $P4$ are not internally consistent either.

In the rest of this paper we disregard these three modes. As their
implementations are so obviously sensitive to the program structure, we do not
see a good way to reconcile them with our semantics-oriented post-processing
specification for aggregation. In fact, in our opinion these modes are best
avoided in high-level logic programs.

\subsection{Lattice-Based Approaches}
\label{sec:lattice-based-approaches}

The remaining three modes, \texttt{min}, \texttt{max} and \texttt{all}, share
one notable property: they are all based on a join-semilattice structure
defined on (subsets of) $U_P$, the set of all ground terms over the alphabet
of $P$.
A \emph{join-semilattice} is a poset $\tuple{S, \leq_S}$ such that
every finite subset $X \subseteq S$ has a least upper bound in $S$,
which we denote, as in the case of complete lattices, $\bigvee X$.
For example, the set of natural numbers with standard order
$\tuple{\mathbb N, \leq}$ is a join-semilattice (with $\bigvee X =
\mathsf{max}\, X$), but it is not a complete lattice. Different modes define the following join-semilattices:

\begin{itemize}
\item \texttt{min} defines the join-semilattice $\tuple{U_P,\leq}$ where $\leq$ is the
  lexicographical ordering on terms (\texttt{=</2}).
  The least upper bound is the minimum.
\item \texttt{max} defines the join-semilattice $\tuple{U_P,\geq}$ where $\geq$ is the
  inverse of $\leq$.
  The least upper bound is the maximum.
\item \texttt{all} defines the join-semilattice
  $\tuple{\powersetfin{U_P},\subseteq}$ where $\powersetfin{U_P}$ is the set of all finite subsets of $U_P$.
  Existing implementations represent sets as lists of terms, which are
  themselves terms.
\end{itemize}

The two additional aggregation approaches, offered by XSB, are also
based on join-semilattices:
\begin{itemize}
\item
XSB generalises the above modes to user-defined join-semilattices with the\\
\texttt{lattice($\bigvee$/3)} mode that is parameterised in a binary
join operator. For instance, we can define the \texttt{min} mode as
\texttt{lattice(min/3)}.
\item
XSB also provides a second user-definable mode \texttt{po($\preceq$/2)} in
terms of a partial order $\preceq$ on $U_P$. This partial order induces a
join-semilattice $\tuple{\powersetfin{U_P},\sqsubseteq}$ where $X \sqsubseteq Y \equiv
\forall x \in X: \exists y \in Y: x \preceq y$.
\end{itemize}

Therefore, in what follows, we only have to deal with lattices that are
essentially subsets of $U_P$, considerably simplifying the formulae.


As we can always reorder arguments and combine multiple join-semilattices into their
product join-semilattice, we assume, without loss of generality, that only the final
argument of a predicate is an output tabling mode.
That is, all ground atoms have the shape $Q(\vect{X},x)$ where $Q$ is
the name of some predicate, $\vect{X}$ is a vector of input arguments
$X_1,X_2,\ldots,X_n$ and $x$ is the value of the output
parameter. We make the simplifying assumption that all predicates are
tabled. If a predicate has only input arguments (like tabling without answer
subsumption), a (constant) dummy output can always be added.
  
Mode-directed tabling groups atoms for a predicate $Q$ by distinct values for
the \emph{input} arguments $\vect{X}$ and aggregates the values of
the \emph{output} argument $x$ into a single term.
Therefore, we model a table of aggregated answers by a function
$table:\tabledomain$ (where $I_P$ is the set of predicate names
in $P$) that maps a pair of a predicate name and inputs to a single aggregated
output.
The set of aggregate answers $U_P^\bot = U_P \cup \{\bot\}$ is the set of
all terms, on which a special element $\bot$ is grafted, to indicate the
lack of an answer.
We extend the chosen order on terms $\preccurlyeq$ such that $\bot$ is an (adjoined) bottom element, that is,
$\forall x \in U_P^\bot:\bot \preccurlyeq x$.
For legibility, we will also sometimes refer to $\tabledomain$ by $\Lcurly$.
The lattice structure $\tuple{U_P^\bot,\preccurlyeq}$ induces a join-semilattice structure
$\tuple{\Lcurly, \preceq}$ on the set of tables, where
$\preceq$ is the pointwise order:
\[
f \preceq g \iff \forall (p,\vect{X}) \in I_P \times U_P^n :
f(p,\vect{X}) \preccurlyeq g(p,\vect{X})
\]
This lattice structure allows us to aggregate over multiple tables, by
aggregating the answers pointwise:
\[
(\bigvee F)(p,\vect{X}) = \bigvee_{f \in F} f(p,\vect{X})
\]

By storing each individual element of $\fix (T_P)$ into a table and then
aggregating over tables we obtain the semantics for mode directed tabling:

Let $\eta:H_P \to (\tabledomain)$ and
$\rho:(\tabledomain) \to \PH$ be defined as:
\begin{align*}
&\eta(p(\vect{X},x))(q,\vect{Y}) =
\begin{cases}
  x    &\text{ if } p = q \wedge \vect{X} = \vect{Y}\\
  \bot &\text{ otherwise}
\end{cases}\\
&\rho(f) = \{p(\vect{X},f(p,\vect{X}))~|~f(p,\vect{X}) \neq \bot\}
\end{align*}
Thus, the function $\eta$ turns an atom into a singleton table, and $\rho$ maps
a table to the set of its true atoms.

To compute the set of all true atoms of a program $P$, we need to
consider the consequence of joining two elements of a semi-lattice in
addition to regular logical consequences. This is because, for
arbitrary lattices, the result of a join can be distinct from any of
its arguments, and thus produce new facts. We define a new
immediate consequence operator~$\T_P$, which extends the regular $T_P$
operator with answers obtained by joins.
Formally, we define $\T_P$ as follows, where $\mathcal P^{\mathrm{fin}}(A)$ denotes the set of all finite subsets of a set $A$:
\begin{equation}
\label{eq:tphat}
\T_P(X) =  \bigcup \{\rho(\bigvee Y)~|~Y \in \mathcal P^{\mathrm{fin}}(\eta(T_P(X)))\}
\end{equation}
One can show that $\T_P$ is continuous, hence monotone. In fact, for linear orders (such as
\texttt{min} and \texttt{max}), $\T_P$ behaves exactly like $T_P$.
Again, we consider the least fixed-point of $\T_P$ to
be the set of all the answers that can be obtained by the logical
rules and the `lattice rules'.

The next step is to discard the
subsumed answers by applying the join operator on the set of
answers. Thus, the set of all true atoms of the program $P$ using
mode-directed tabling is given by:
\begin{equation}
\label{eq:tphatsemantics}
\rho\left(\textstyle\bigvee_{x \in \fix (\T_P)} \eta(x)\right)
\end{equation}
Obviously, when $\fix (\T_P)$ is infinite, the least upper bound above
does not necessarily exist. That is why to give a full denotational
semantics of answer subsumption in the next subsection, we model
tables in a more abstract way as complete lattices. Now, to provide
some intuition, we give an example in which the least upper bound
exists.

\begin{example}
  \label{ex:shortest-path}
  Consider the example from the introduction, rewritten using XSB's
  lattice answer subsumption for the sake of variety:
  \begin{CodeVerbatim}
  :- table p(lattice(_,_,min/3)).
  :- table e/3.
  p(X,Y,1) :- e(X,Y,nt).
  p(X,Y,D) :- p(X,Z,D1), p(Z,Y,D2), D is D1 + D2
  e(a,b,nt). e(b,c,nt). e(a,c,nt).
  min(X,Y,Z) :- Z is min(X,Y).
  \end{CodeVerbatim}
  Note that we have additionally tabled \texttt{e/3} and added a dummy output
  parameter (\texttt{nt} stands for not tabled), as described above.
  The don't cares (\texttt{\_}) in the tabling directive indicate that they
  are not part of the lattice.
  In Yap's terminology: they use the \texttt{index} tabling mode.
  The least fixed-point semantics of this program, that is $\fix (\T_P)$, is
  given by the following set:
  \[
  \{\texttt{e(a,b,nt)},\texttt{e(b,c,nt)},\texttt{e(a,c,nt)},
  \texttt{p(a,b,1)},\texttt{p(b,c,1)},\texttt{p(a,c,1)},\texttt{p(a,c,2)}\}
  \]
  The complete lattice on the final argument of \texttt{p} is
  $\tuple{\mathbb{N},\geq}$, the reversed standard order.
  The least upper bound in this lattice is the usual \emph{infimum} on
  natural numbers.

  Interpreted by this lattice, the semantics is
  \begin{align*}
    &\rho\left(\textstyle\bigvee_{x \in \fix (\T_P)} \eta(x)\right)\\
    =\ &\rho(\bigvee \{\eta(\texttt{e(a,b,nt)}),\eta(\texttt{e(b,c,nt)}),
    \eta(\texttt{e(a,c,nt)}),\\
    & \hspace{2.5em} \eta(\texttt{p(a,b,1)}), \eta(\texttt{p(b,c,1)}),
    \eta(\texttt{p(a,c,1)}),\eta(\texttt{p(a,c,2)})\})\\
    =\ &\rho(t) \text{ where } t(q,x,y) =
    \begin{cases}
      \mathtt{nt}           & \text{ if } q = \texttt{e}\\
      1            & \text{ if } (x = \texttt{a} \wedge y = \texttt{b})
                      \vee (x = \texttt{b} \wedge y = \texttt{c})\\
      \min \{1,2\} & \text{ if } x = \texttt{a} \wedge y = \texttt{c}\\
      \bot         & \text{ otherwise}
    \end{cases}\\
    =\ &\{\texttt{e(a,b,nt)},\texttt{e(b,c,nt)},\texttt{e(a,c,nt)},
    \texttt{p(a,b,1)},\texttt{p(b,c,1)},\texttt{p(a,c,1)}\}
  \end{align*}
  Only the atoms representing the shortest paths are retained, as expected.
\end{example}

\begin{example}
This example illustrates why we need to extend the $T_P$ operator to
include the results of the lattice operations, that is, why we need
the $\T_P$ operator. Consider the lattice $\{a,b,c,d\}$, with $a,b \leq
c$ and $c \leq d$, which we use in the following program:

\begin{CodeVerbatim}
   lub(a,b,c). lub(a,c,c). lub(a,d,d).
   lub(b,a,c). lub(b,c,c). lub(b,d,d).
   lub(c,d,d).
   lub(X,X,X).   

   :- table p(lattice(lub/3)).

   p(a).
   p(b).
   p(d) :- p(c).
\end{CodeVerbatim}

\noindent
The regular immediate consequence gives us $\fix(T_P) = \{
\mathtt{p(a)}, \mathtt{p(b)} \}$, which means that $\rho(\bigvee
\eta(\fix(T_P))) = \{ \mathtt{p(c)} \}$. The atom $\mathtt{p(c)}$ does
not follow from the logical inference, but from the lattice's join
operator. It is included in the overall answer thanks to the
post-processing step, but its logical consequences are not. With the
$\T_P$ operator we have $\fix(\T_P) = \{ \mathtt{p(a)}, \mathtt{p(b)},
\mathtt{p(c)}, \mathtt{p(d)} \}$, and so $\rho(\bigvee
\eta(\fix(\T_P))) = \{ \mathtt{p(d)} \}$, which is the intended
semantics.
\end{example}

\subsection{Answer Subsumption for Arbitrary Lattices}
\label{sec:tablesWithArbitraryLattices}

\newcommand{\toLattice}[1]{\lceil {#1} \rceil}
\newcommand{\fromLattice}[1]{\lfloor {#1} \rfloor}

Even though at any point of computation each table is finite, it is
potentially infinite when a program produces infinitely many
answers. Thus, to give a denotational semantics for answer
subsumption, a join-semilattice on terms is not enough, as we need
least upper bounds of infinite sets, i.e. a complete lattice
structure. For example, the most natural candidate for the types of
values in the case of the \texttt{all} mode is $\tuple{\powerset{U_P},
  \subseteq}$, the complete lattice of all subsets of $U_P$,
which cannot be modelled by (finite) terms. In general, every
semilattice can be extended to a complete lattice via
MacNeille~\citeyear{posets} completion.

Thus, we do not impose any order on the set $U_P^\bot$, and the type
of the table becomes $(I_P \times U^n_P \to L)$ for a complete lattice
$L$. For each predicate $p \in I_P$, we also need two
bottom-preserving \emph{abstraction} and \emph{representation}
functions: $\toLattice{\text{-}}_p: U_P^\bot \to L$ and
$\fromLattice{\text{-}}_p : L \to U_P^\bot$ respectively. We require
$\toLattice{\text{-}}_p$ to be a retraction of
$\fromLattice{\text{-}}_p$, that is, $\toLattice{\fromLattice{x}_p}_p
= x$. Since we want the two functions to preserve bottoms, the least
element of $L$ denotes `no value'.
With this, we give new definitions of $\eta$ and $\rho$, appropriately
adding abstraction and representation, where $\bot^{\!L}$ is the least
element of $L$:
\begin{align*}
&\eta(p(\vect{X},x))(q,\vect{Y}) =
\begin{cases}
  \toLattice{x}_p    &\text{ if } p = q \wedge \vect{X} = \vect{Y}\\
  \bot &\text{ otherwise}
\end{cases}\\
&\rho(f) = \{p(\vect{X}, \fromLattice{f(p,\vect{X})}_p)~|~f(p,\vect{X}) \neq \bot^{\!L}\}
\end{align*}

To give the semantics, we define the $\T_P$ operator exactly as
in~\eqref{eq:tphat} but using the new definitions of $\eta$ and
$\rho$. It is easy to see that it is monotone, so it always has a
least fixed point. The semantics of the entire program is given again
as in~\eqref{eq:tphatsemantics}.

\subsection{Lattice Semantics for Stratified Programs}
\label{sec:latticeStratified}

For general programs, we use stratification to distinguish between
predicates that imply and are implied by tabled values.  We define the
\emph{depends on} relation $\ltimes$ as follows: for any two
predicates $\mathtt{p}$ and $\mathtt{q}$, it is the case that
$\mathtt{p} \ltimes \mathtt{q}$ if and only if there exists a clause
$\texttt{p(...):-...,q(...),...}$.  We say that $\mathtt{p}$ and
$\mathtt{q}$ are in the same stratum if $\mathtt{p} \ltimes^{\!{+}}
\mathtt{q}$ and $\mathtt{q}\ltimes^{\!{+}}\mathtt{p}$, where
$\ltimes^{\!{+}}$ is the reflexive and transitive closure of
$\ltimes$.  Put differently, a stratum is a strongly connected
component of the dependency graph defined by $\ltimes$.  The relation
$\ltimes$ induces a partial ordering on the set of all strata $S$,
that is, for $X,Y \in S$, it is the case that $X \leq Y$ if and only
if there exists $\texttt{p} \in X$ and $\texttt{q} \in Y$ such that
$\mathtt{p}\ltimes^{\!{+}}\mathtt{q}$.

A stratum forms a logical unit to which the least fixed point
semantics and aggregation are applied in turn: For each stratum $X \in
S$, we can define its semantics $M_X \subseteq \PH$ as follows: $M_X =
\rho(\bigvee \eta( \lfp(\T_Q)))$, where $Q = P_X \cup \bigcup_{Y <
  X}M_Y$ and $P_X$ is the set $\mathsf{ground}(P)$ restricted to the
predicates in the stratum~$X$, while $\bigcup_{Y < X}M_Y$ should be
understood as a set of facts.  Informally, this means that to give a
semantics for a stratum, we first compute the semantics of the strata
below, use the results as a set of facts added to the part of the
program in the current stratum, compute the fixed point, and finally
perform the aggregation step using the join operator.  There are
always finitely many strata, so $M_X$ is well-defined.
The semantics of the program $P$ is then the aggregation of the sum of
the interpretations of all the strata, that is, $\bigcup_{X \in S}M_X$.

Stratification ensures that the answers for a predicate are always
aggregated before they are used by another predicate, unless there is a cyclic
dependency between them.
%
for example,
consider the following variation on the shortest path program:
\begin{CodeVerbatim}
   :- table p(index,index,min).
   e(1,2). e(2,3). e(1,3).
   p(X,Y,1) :- e(X,Y)
   p(X,Y,D) :- p(X,Z,D1),p(Z,Y,D2), D is D1 + D2.
   s(X,Y,D) :- p(X,Y,D).
\end{CodeVerbatim}

\noindent
Without stratification, the semantics is given by
$\rho(\bigvee\eta(\fix(\T_P))$ which contains \texttt{s(1,3,2)}, as a
consequence of \texttt{p(1,3,2)} before aggregation.
However, if we stratify the program as discussed above, the rules
for \texttt{p} end up in a lower stratum than \texttt{s}. Therefore,
the results for \texttt{p} will be aggregated by \texttt{min}, before
any consequence is derived from it.
Since \texttt{p(1,3,2)} is subsumed, \texttt{s(1,3,2)} is no
longer derived.

When two predicates are interdependent (and therefore in the same stratum), but
only one of them is tabled, the answers for the untabled predicate are not
subsumed. Stratification then gives a result that appears inconsistent:
\begin{CodeVerbatim}
   :- table even(min).

   even(0).
   even(X) :- odd(Y), Y is X - 1.
   odd(X) :- even(Y), Y is X - 1.
\end{CodeVerbatim}
Our semantics interprets this program as the set
$$\rho(\bigvee \eta(\fix(\T_P))) = \{ \mathtt{even(0)}, \mathtt{odd(1)}, \mathtt{odd(3)}, \mathtt{odd(5)}, \mathtt{odd(7)}, \ldots \}.$$
While $\{ \mathtt{even(0)}, \mathtt{odd(1)} \}$ seems equally reasonable.
Because we treat subsumption as a post-processing step \emph{per stratum},
which means that inter-dependent predicates are resolved as if no answers were
subsumed.
Subsumption only affects predicates in the strata above.
For instance, assume we add the following (non-tabled) predicate to the program:
\begin{CodeVerbatim}
   also_odd(X) :-  even(Y), Y is X - 1.
\end{CodeVerbatim}
It is in a different stratum than $\mathtt{even}$ and $\mathtt{odd}$,
so its semantics depends on the semantics of $\mathtt{even}$ after the
subsumption step. This means that the semantics together with the
$\mathtt{also\_odd}$ predicate is given as:
$$\{ \mathtt{even(0)}, \mathtt{also\_odd(1)}, \mathtt{odd(1)}, \mathtt{odd(3)}, \mathtt{odd(5)}, \mathtt{odd(7)}, \ldots \},$$
Here \texttt{also\_odd} behaves like the alternative suggested for \texttt{odd}
above.
Importantly, programs like the one above do not satisfy our
correctness condition for the greedy strategy given in
Section~\ref{sec:fusion}.


%
%
%
%

\section{Generalised Answer Subsumption Semantics}
\label{sec:fusion}

The previous section specifies the semantics of tabling with lattice-based
answer subsumption in terms of a post-processing aggregation. However, the
existing implementations do not actually first compute the least Herbrand
model. Instead, they greedily aggregate intermediate results during SLD-resolution.  This
makes it feasible to, for instance, compute the shortest path in a cyclic graph
in a finite amount of time, as well as generally improving efficiency.
However, as the examples in the introduction acutely demonstrate, this
greedy strategy is not always valid.
This section characterises the greedy strategy as a form of generalised
semantics for logic programs and considers its correctness with respect to our
postulated specification.


\subsection{Generalised Semantics}

Again, we assume that we work in a single stratum. We can
capture alternate greedy strategies as \emph{generalised semantics} of $P$ in 
terms of structures $\tuple{L,\eta^L,T^L_P}$, where:
%
\begin{enumerate}
\item $L$ is a complete lattice,
\item $\eta^L : H_P \to L$ is a function that `embeds' terms in the lattice,
\item $T^L_P : L \to L$ (a \emph{generalised} immediate consequence operator) is monotone,
\item $L$ is generated by $\eta^L (H_P)$, which means that for every $x \in L$, there exists an
  $X \subseteq \eta^L(H_P)$ such that $x = \bigvee X$.
\end{enumerate}

Note that in general we do not need a counterpart of the $\rho$
function: in the post-processing semantics given in
Section~\ref{sec:denotationalSemanticsAnswerSubsumption}, we need $\rho$
to define the $\T_P$ operator and move between strata. Here, we use
this semantics to capture correctness within a single stratum, and the
immediate consequence operator is not defined, but it is given. This
allows us to generalise the whole table to a lattice, which simplifies
the generalised semantics.

The generalised semantics of the program $P$ is given by
$\fix(T^L_P)$, which exists due to the Knaster--Tarski theorem.  It is
easy to verify that the regular least fixed-point semantics is a valid
instance of this generalised semantics.
Also note that the generalised semantics does not depend on $\eta^L$ or the
fact that $L$ is generated by $\eta^L$. We need these facts in a
moment to establish a correctness criterion.

One obvious instantiation of this semantics is with the $\T_P$
operator defined in Section~\ref{sec:tablesWithArbitraryLattices},
where $\iota : X \rightarrow \powerset{X}$ is the singleton function:
\begin{equation*}
\tuple{\PH, \iota, \T_P}
\end{equation*}

We say that an instance of the generalised semantics is a correct
implementation strategy only if yields the same result as the
post-processing semantics defined in
Section~\ref{sec:denotationalSemanticsAnswerSubsumption}.
More formally, a generalised semantics $\tuple{L,\eta^L,T^L_P}$ is
\emph{correct} if and only if
\begin{equation}\label{eq:correctness}
\fix(T^L_P) = \interp{\eta^L}(\fix(\T_P)),
\end{equation}
where, for convenience, we define, for any
lattice $L$, set $S$, and function $f : S \to L$, the function
$\interp f : \powerset{S} \to L$ as follows:
\begin{equation*}
\interp{f}(Y) = \bigvee_{x \in X} f(x)
\end{equation*}
Intuitively, it means that evaluating the whole program with no answer
subsumption and then obtaining the final result using $L$'s join operation on
the answers is the same as computing every step with the lattice $L$ (which is
usually more efficient).
The following theorem gives sufficient conditions for
the equation~\eqref{eq:correctness} to hold.

\begin{thm}[Fixed-Point Fusion \cite{DBLP:conf/acmmpc/Backhouse00}]\label{thm:backhouse}
  Let $\tuple{X, \leq_X}$ and $\tuple{Y, \leq_Y}$ be posets.
  Let $f: X \to X$ and $g : Y \to Y$ be two functions with least fixed points.
  Let $h : X \to Y$ be a function. Assume the following:
\begin{enumerate}[itemsep=1ex,leftmargin=1cm,label=(\alph*)]
\item It is the case that $h \circ f = g \circ h$.
\item The function $h$ has an upper Galois adjoint, that is, there exists a
  function $j : Y \to X$ such that
  $h(x) \leq_Y y \Longleftrightarrow x \leq_X j(y)$ for all $x \in X$ and
  $y \in Y$.
\end{enumerate}
Then, $\fix (g) = h (\fix (f))$.
\end{thm}

The equation~\eqref{eq:correctness} is clearly an instance of this theorem's
conclusion, with $f = \T_P$, $g = T^L_P$ and $h = \interp{\eta^L}$.
Yet, the theorem only applies if we can show that
(a) $\interp{\eta^L} \circ \T_P = T^L_P \circ \interp{\eta^L}$, and that 
(b) $\interp{\eta^L}$ has an upper Galois adjoint. 
Fortunately, (b) readily follows from the fact that $\interp{\eta^L}$ is
continuous, and that every continuous function has an upper Galois
adjoint~\cite{DBLP:conf/acmmpc/Backhouse00}.

Intuitively the remaining condition (a) means that $\interp{\eta^L}$
should preserve immediate consequences.
In other words, subsumption of a conventional immediate consequence should be
the generalised immediate consequence of $\interp{\eta^L}$.

\subsection{Greedy Strategy}
\label{sec:greedy-strategy}

For better performance, practical implementations of tabling with
answer subsumption use a greedy strategy, which means that they
remove subsumed answers as soon as possible, and not as a single
post-processing step. We can try to express it as the following instance
of the generalised semantics, in terms of the functions defined
in Section~\ref{sec:tablesWithArbitraryLattices}:
\begin{equation*}
\tuple{I_P \times U^n_P \to L, \eta, \interp{\eta} \circ T_P \circ \rho}
\end{equation*}
The function $\rho$ extracts the set of true ground atoms from a table,
which is a subset of the Herbrand base $H_P$, so we can indeed express the
`greedy' immediate consequence operator in terms
of the standard immediate consequence operator $T_P$, subsumption
$\interp{\eta}$ and extraction $\rho$.
In fact this definition is quite convenient, because tabled Prolog systems
readily provide an efficient implementation of $T_P$.

The question remains if the function $\interp{\eta} \circ T_P \circ
\rho$ has a fixed point.  Luckily, given a tuple
$\tuple{L,\eta^L,T^L_P}$, where $L$ is generated by $\eta^L (H_P)$,
the condition (a) is enough for $T^L_P$ to be monotone (this is where
the assumption that $\eta^L(H_P)$ generates $L$ comes in handy):
\begin{thm} \label{thm:monotone}
  Let $\tuple{L,\eta^L,T^L_P}$ be as above.
  If $\interp{\eta^L} \circ \T_P = T^L_P \circ \interp{\eta^L}$ (the
  condition (a)), then $T^L_P$ is~monotone.
\end{thm}
%
%
Additionally, it is the case that $\interp{\eta} \circ T_P = \interp{\eta} \circ \T_P$.
Thus, to show correctness of a program under the greedy semantics,
we need only show the following:
\begin{equation}\label{eq:specific-correctness}
  \interp{\eta} \circ T_P = \interp{\eta} \circ T_P \circ \rho \circ \interp{\eta}
\end{equation}
\begin{example}\label{ex:greedy-bad}
Reconsider the invalid program from the introduction.
We show that the condition~\ref{eq:specific-correctness} does not hold for this
program under semantics $\tuple{\Lcurly,\preceq,\eta,T_P^\Lcurly}$ where
$H_P = \{\texttt{p(0)},\texttt{p(1)},\texttt{p(2)},\texttt{p(3)}\}$ and
$(\preccurlyeq) = (\leq)$ is the partial order on terms.
By means of the following counter example: A simple calculation shows that
$(\interp{\eta} \circ T_P)\big(\{\texttt{p(0)},\texttt{p(1)}\}\big)
\neq
(\interp{\eta} \circ T_P \circ \rho \circ \interp{\eta})\big(
\{\texttt{p(0)},\texttt{p(1)}\})\big)$.

This example also explains the odd behaviour of XSB, Yap and \BProlog{}:
\texttt{p(0)} is subsumed by \texttt{p(1)}, therefore the body of the third
rule in the program cannot be satisfied and \texttt{p(3)} (the correct answer)
is never produced.

Interestingly, when we change the last rule to \texttt{p(3) :- p(0)}, the
result of the query \texttt{?-p(X).} changes once more in all systems,
although logically both rules should behave identically.
Furthermore, different implementations handle this rule differently.
For instance, XSB reasonably disallows calls where \texttt{lattice}-mode
arguments are not free, and the latter rule therefore produces an error
message. While Yap instead produces \texttt{X=3}, because its
batched-mode evaluation immediately derives \texttt{p(1)} from the
fact \texttt{p(1)}.
\end{example}

\begin{example}
  Now, consider the shortest path program from Example~\ref{ex:shortest-path}
  under semantics $\tuple{\Lcurly,\preceq,\eta,T_P^\Lcurly}$ and
  $(\preccurlyeq) = (\geq)$.
  We prove the correctness condition by proving that $\mathit{lhs} \preceq \mathit{rhs}$
  and $\mathit{lhs} \succeq \mathit{rhs}$.
  Then by anti-symmetry of $\preceq$, the correctness condition holds.

  \noindent
  \textbf{($\succeq$-direction)} Because $\interp{\eta} \circ T_P$ is
  monotone, and for this case $\rho \circ \interp{\eta}$ is deflative, i.e.
  $\rho(\interp{\eta}(X)) \subseteq X$ for all $X \subseteq H_P$, we are done
  with this direction.
  %
  
  \noindent
  \textbf{($\preceq$-direction)}
  A \texttt{p(x,y,d)}-atom is either introduced by an edge \texttt{e(x,y,nt)}
  or by two other atoms \texttt{p(x,z,d1)} and \texttt{p(z,y,d2)}.
  In the former case $d = 1$, in the latter case $d = d_1 + d_2$.
  It is easy to see that if an atom \texttt{p(x,y,1)} is introduced by $T_P$,
  it is also introduced by $T_P \circ \rho \circ \interp{\eta}$.
  In the latter case, $d_1$ and $d_2$ are at least as large as their
  corresponding entries in the table produced by $\interp{\eta}$.
  Hence $d$ must be at least as large as any corresponding distance in
  the set produced by $T_P \circ \rho \circ \interp{\eta}$.
  Hence the infimum of $d$s produced by $T_P$ must be at least as large as
  any corresponding distance. Finally this means that the entry in the
  table produced by $\interp{\eta} \circ T_P$ must be at least as large
  as the entry in the table produced by
  $\interp{\eta} \circ T_P \circ \rho \circ \interp{\eta}$.
  
\end{example}



\section{Related Work}
\label{sec:relatedWork}

As far as we know, ``output'' subsumption for tabling was first proposed by Van
Hentenryck et al.~\shortcite{VanHentenryck:1993:AIP:864486} in the context of
abstract interpretation.

\paragraph{Tabling Modes for Dynamic Programming}
In dynamic programming, an optimal solution to a problem is defined in terms
of the optimal solutions of smaller sub-problems.
This intuition is in fact captured by the correctness condition
(Equation~\ref{eq:specific-correctness}): it states that the solution found
by examining all sub-solutions, is equal to examining only the optimal
solutions. This is good news, because it implies correctness for
dynamic programming.

We have already discussed the tabling modes of Yap~\cite{santos2013mode} and
XSB~\cite{swift2012} at length in Section~\ref{sec:mode-directed-tabling}.
XSB's lattice based answer subsumption is more suitable for implementing
techniques that require more general lattices than simple minimum and
maximum, such as abstract interpretation.

Guo and Gupta~\shortcite{guo2004mode,guo2008modeExpanded} implemented 5 tabling
modes in ALS-Prolog with the aim of simplifying and accelerating
dynamic programming. 
These modes are \texttt{+} (indexed), \texttt{-−}(only the first
answer is retained), \texttt{min} (minimum), \texttt{max} (maximum) and
\texttt{last} (the last answer is retained).
The correspondence to Yap's tabling modes is obvious, but there is no
equivalent for Yap's \texttt{all} and \texttt{sum} modes.
Answers are grouped by distinct values for the \texttt{+} arguments
and the remaining arguments are aggregated based on their mode.

Zhou~\etal~\shortcite{zhou2010mode} added tabling to \BProlog{} with the same
purpose in mind.
Consequently, the tabling modes they support are identical, except they do not
implement a \texttt{last} mode.
Instead they support cardinality constraints, which limit the answers that
are stored in the table to the first $N$ optimal answers for some positive
integer $N$.
Also supported is an \texttt{nt} (not-tabled) tabling mode, which is used to
pass around global constants efficiently.
From the perspective of the tabling system \texttt{nt} arguments do not exist,
and thus are never stored in the table.
\vspace{-4mm}
\paragraph{Haskell}
Vandenbroucke\etal~\shortcite{fixingNonDeterminism} have added
lattice answer subsumption to their tabling implementation in Haskell. It is
based on the effect handlers approach.
\vspace{-4mm}
\paragraph{Abstract interpretation} Our approach bears a strong resemblance to
abstract interpretation~\cite{Cousot1992,abramskiAi}.
Unlike answer subsumption, abstract interpretation admits approximate
solutions, implying a weaker correctness condition where equality is
replaced by an order relation.
\vspace{-4mm}
\paragraph{Matroids and Greedoids}
Other set theoretic structures (besides lattices) such as \emph{matroids}
~\cite{matroid} and \emph{greedoids}~\cite{korte1991greedoids}, have been
developed to analyse greedy algorithms and show their optimality.
As answer subsumption is essentially a greedy strategy, we plan to re-examine
answer subsumption in this new context in the future.

\section{Conclusion and Future Work}
\label{sec:conclusion}

In many instances of tabling only the optimal answers to a query are relevant.
To improve performance over a naive generate-and-aggregate approach,
various forms of answer subsumption that greedily combine these answers have
been developed in the literature.
However, their semantics has never been described formally.
An operational understanding is always an option in this case, and although
often useful, it is a far cry from the declarative ideal that tabling promises.

We define a high-level semantics for answer subsumption based on lattice
theory. Then we generalise it to establish a correctness condition indicating
when it is safe to use (greedy) answer subsumption.
We show several examples where the existing implementations of answer
subsumption fail that condition and derive an erroneous result.

This condition is sufficient, but not necessary: there may still exist programs
that do not satisfy the condition, for which the greedy strategy nevertheless
delivers correct results.
Since we have not run across any non-contrived examples of such programs, we
contend that this apparent lack of necessity is an artefact of our rather
coarse semantics, which we intend to refine in future work.

The verification of correctness does constitute a non-trivial effort.
Hence, manually proving the correctness condition for realistically sized
programs could be unfeasible in practice.
Ideally we would have an automated analysis that warns the programmer if it
fails to establish the correctness condition.
This is future work.

\paragraph*{Acknowledgements}
  We would like to thank Bart Demoen for enlightening discussions during the preparation of this paper.

\bibliographystyle{acmtrans}
\bibliography{bibliographyDatabase}

\appendix
\section{Proofs and Calcuclations}
\label{sec:proofs}

\subsection{$\omega$-continuity of the $\T_P$ operator}

Let $D = \{ d_0, d_1, d_2, \ldots \}$ be an $\omega$-chain, that is, $d_0 \subseteq d_1 \subseteq d_2 \subseteq \ldots$. We need to show that $\T_P(\bigcup D) = \bigcup \T_P(D)$. The left-hand side:
\begin{align}
\T_P(\bigcup D) &= \bigcup \{\rho(\bigvee Y)~|~Y \in \mathcal P^{\mathrm{fin}}(\eta(T_P(\bigcup D)))\} \tag{def. of $\T_P$}
\\
&=
\bigcup \{\rho(\bigvee Y)~|~Y \in \mathcal P^{\mathrm{fin}}(\eta(\bigcup T_P(D)))\} \tag{$T_P$ is $\omega$-cont.}
\\
&= \{ x ~|~ \exists Y \in \mathcal P^{\mathrm{fin}}(\eta(\bigcup T_P(D))).\ x \in \rho(\bigvee Y)\} \tag{def. of $\bigcup$}
\end{align}

\noindent
The right-hand side:
\begin{align}
\bigcup \T_P(D) &=
\bigcup \{ \T_P(d) ~|~ d \in D\} \tag{image}
\\
&= \bigcup \{ \bigcup \{\rho(\bigvee Y)~|~ Y \in \mathcal P^{\mathrm{fin}}(\eta(T_P(d)))\} ~|~ d \in D\} \tag{def. of $\T_P$}
\\
&= \{ x ~|~ \exists d \in D.\ \exists Y \in \mathcal P^{\mathrm{fin}}(\eta(T_P(d))).\ x \in \rho(\bigvee Y) \} \tag{def of $\bigcup$} 
\end{align}

\noindent
Thus, it is enough to show that $\exists d \in D.\ \exists Y \in \mathcal P^{\mathrm{fin}}(\eta(T_P(d))).\ x \in \rho(\bigvee Y)$ if and only if $\exists Y \in \mathcal P^{\mathrm{fin}}(\eta(\bigcup T_P(D))).\ x \in \rho(\bigvee Y)$. The $(\Rightarrow)$ direction is trivial. For the $(\Leftarrow)$ direction pick $d_N$ with the lowest $n \in \mathbb{N}$ such that $Y \subseteq \eta(T_P(d_n))$, which exists, since $Y$ is finite.

\subsection{Proof of Theorem~\ref{thm:monotone}}

Let $x \leq_L y$. Since $L$ is generated by $\eta^L(H_P)$, there exist
$X \subseteq Y \subseteq H_P$ such that
$x = \bigvee \eta^L(X)$ and $y = \bigvee \eta^L(Y)$.
\begin{align*}
T^L_P(x) \makebox[2.5em]{\centering =} &  T^L_P(\bigvee \eta^L(X))
\\
\makebox[2.5em]{\centering =} &
T^L_P(\interp{\eta^L}(X)) \tag{def. of $\interp{\eta^L}$}
\\
\makebox[2.5em]{\centering =} &
\interp{\eta^L}(\T_P(X)) \tag{assumption}
\\
\makebox[2.5em]{\centering $\leq_L$} &
\interp{\eta^L}(\T_P(Y)) \tag{composition preserves monotonicity}
\\
\makebox[2.5em]{\centering =} &
T^L_P(\interp{\eta^L}(Y)) \tag{assumption}
\\
\makebox[2.5em]{\centering =} &
T^L_P(\bigvee \eta^L(Y)) \tag{def. of $\interp{\eta^L}$}
\\
\makebox[2.5em]{\centering =} &
T^L_P(y)
\end{align*}
\subsection{Counter Example for Example~\ref{ex:greedy-bad}}

\begin{align*}
  (\interp{\eta} \circ \T_P)\big(\{\texttt{p(0)},\texttt{p(1)}\}\big)
  &= \interp{\eta}\big(\{\texttt{p(0)},
  \texttt{p(1)},
  \texttt{p(2)},
  \texttt{p(3)}\}\big)\\
  &= t \text{ such that  $t(Q) =$ if $Q = \texttt{p}$ then 3 else $\bot$}\\
  &\neq u \text{ such that  $u(Q) =$ if $Q = \texttt{p}$ then 2 else $\bot$}\\
  &= \interp{\eta}\big(\{\texttt{p(2)}\})\\
  &= (\interp{\eta} \circ \T_P)\big(\{\texttt{p(1)}\}\big)\\
  &= (\interp{\eta} \circ \T_P \circ \rho)(t) \text{ such that $t(Q) =$ if
    $Q = \texttt{p}$ then 1 else $\bot$}\\
  &= (\interp{\eta} \circ \T_P \circ \rho \circ \interp{\eta})\big(
  \{\texttt{p(0)},\texttt{p(1)}\})\big)
\end{align*}

\section{Additional Examples}
\label{app:examples}

This appendix contains some additional examples that do not fit in the
main part of the paper because of the page limit, but which could be
useful in understanding the details of our semantics for tabling with
answer subsumption.

\subsection{The Extended Immediate Consequence Operator}

This example illustrates why we need to extend the $T_P$ operator to
include the results of the lattice operations, that is, why we need
the $\T_P$ operator. Consider the lattice $\{a,b,c,d\}$, with $a,b \leq
c$ and $c \leq d$, which we use in the following program:

\begin{CodeVerbatim}
   lub(a,b,c). lub(a,c,c). lub(a,d,d).
   lub(b,a,c). lub(b,c,c). lub(b,d,d).
   lub(c,d,d).
   lub(X,X,X).   

   :- table p(lattice(lub/3)).

   p(a).
   p(b).
   p(d) :- p(c).
\end{CodeVerbatim}

\noindent
The regular immediate consequence gives us $\fix(T_P) = \{
\mathtt{p(a)}, \mathtt{p(b)} \}$, which means that $\rho(\bigvee
\eta(\fix(T_P))) = \{ \mathtt{p(c)} \}$. The atom $\mathtt{p(c)}$ does
not follow from the logical inference, but from the lattice's join
operator. It is included in the overall answer thanks to the
post-processing step, but its logical consequences are not. With the
$\T_P$ operator we have $\fix(\T_P) = \{ \mathtt{p(a)}, \mathtt{p(b)},
\mathtt{p(c)}, \mathtt{p(d)} \}$, and so $\rho(\bigvee
\eta(\fix(\T_P))) = \{ \mathtt{p(d)} \}$, which is the intended
semantics.

\subsection{Circular Dependencies}
\label{sec:circularDeps}

The following example shows what happens when two predicates are
interdependent, but only one of them is tabled:

\begin{CodeVerbatim}
   :- table even(min).

   even(0).
   even(X) :- odd(Y), Y is X - 1.
   odd(X) :- even(Y), Y is X - 1.
\end{CodeVerbatim}

\noindent
Our semantics interprets this program as the set
$$\rho(\bigvee \eta(\fix(\T_P))) = \{ \mathtt{even(0)}, \mathtt{odd(1)}, \mathtt{odd(3)}, \mathtt{odd(5)}, \mathtt{odd(7)}, \ldots \}.$$
One other possible candidate would be $\{ \mathtt{even(0)},
\mathtt{odd(1)} \}$. It is because we treat subsumption as a
post-processing step \emph{per stratum}, which means that
inter-dependent predicates are resolved as if no answers were
subsumed. Subsumption affects predicates in the strata above. For
instance, assume we add the following (non-tabled) predicate to the
program:

\begin{CodeVerbatim}
   also_odd(X) :-  even(Y), Y is X - 1.
\end{CodeVerbatim}

\noindent
It is in a different stratum than $\mathtt{even}$ and $\mathtt{odd}$,
so its semantics depends on the semantics of $\mathtt{even}$ after the
subsumption step. This means that the semantics together with the
$\mathtt{also\_odd}$ predicate is given as:
$$\{ \mathtt{even(0)}, \mathtt{also\_odd(1)}, \mathtt{odd(1)}, \mathtt{odd(3)}, \mathtt{odd(5)}, \mathtt{odd(7)}, \ldots \}$$

Importantly, programs like the one above do not satisfy our
correctness condition for the greedy strategy.

\subsection{Example of answer subsumption for arbitrary lattices}
Consider the shortest path program, now interpreted in the lattice
$\tuple{\mathbb{N}\cup\{\infty\},\leq_{\infty}}$, the natural numbers
extended with infinity with the canonical order, and abstractions and
representations
$\toLattice{\cdot}^{\infty}_e$, $\fromLattice{\cdot}^{\infty}_e$,
$\toLattice{\cdot}^{\infty}_p$ and $\fromLattice{\cdot}^{\infty}_p$, where:
\begin{align*}
  &\toLattice{\texttt{nt}}^{\infty}_e = 1 &&\fromLattice{x}^{\infty}_e = \texttt{nt}\\
  &\toLattice{\texttt{d}}^{\infty}_p  = \begin{cases}
    \infty & \text{if } \texttt{d} = \texttt{infty}\\
    d      & \text{otherwise}
  \end{cases}
  &&\fromLattice{d}^{\infty}_p = \begin{cases}
    \texttt{d} &\text{if } d \in \mathbb{N}\\
    \texttt{infty} & \text{otherwise}
  \end{cases}
\end{align*}
This lattice computes the \emph{longest} path in the graph, demonstrating
that a change in the interpreting lattice can change the result of the
program entirely.
Now suppose the edge \texttt{e(c,a,nt)} is added to the program, creating a
cycle.
The least fixed-point semantics $\fix (\T_P)$ becomes infinite, and
$\rho(\bigvee_{x \in \fix (\T_P)} \eta^{\infty}(x))$ contains only paths of
infinite length, represented by atoms such as \texttt{p(a,c,infty)}.
Although such least fixed-points are not constructively computable in
practice, from a theoretical point of view they demonstrate the essence
of this approach quite well.

\label{lastpage}
\end{document}
